Large Resistance Change on Magnetic Tunnel Junction based Molecular Spintronics Devices

Pawan Tyagi[1,2*], and Edward Friebe[1],

*University of the District of Columbia, Department of Mechanical Engineering, 4200 Connecticut Avenue NW Washington DC-20008, USA[1]*
*Email\*: ptyagi@udc.edu*
*University of Kentucky, Chemical and Materials Engineering Department, 177 F Paul Anderson Hall, Lexington, KY-40506, USA[2]*

**Abstract:** Molecular bridges covalently bonded to two ferromagnetic electrodes can transform ferromagnetic materials and produce intriguing spin transport characteristics. This paper discusses the impact of molecule induced strong coupling on the spin transport. To study molecular coupling effect organometallic molecular complex (OMC) was bridged between two ferromagnetic electrodes of a magnetic tunnel junction (Ta/Co/NiFe/AlOx/NiFe/Ta) along the exposed side edges. OMCs induced strong iter-ferromagnetic electrode coupling to yield drastic changes in transport properties of the magnetic tunnel junction testbed at the room temperature. These OMCs also transformed the magnetic properties of magnetic tunnel junctions. SQUID and ferromagnetic resonance studies provided insightful data to explain transport studies on the magnetic tunnel junction based molecular spintronics devices.

**Introduction:** Connecting magnetic molecules between two ferromagnetic electrodes opens flood gate of opportunities for observing new phenomenon and making novel devices [1, 2]. Initial studies focused on sandwiching molecules between two ferromagnetic leads [3]. In a more popular approach molecules were placed in a break-junction on a nanowire on a planar insulating substrate [4]. However, these two approaches have been extremely difficult to mass produce robust molecular spintronics devices [5]. With conventional approaches it is also almost impossible to conduct extensive magnetic studies to explore the true effect of molecules on the magnetic properties of the molecular spintronics devices [6]. To date most of the experimental studies have only focused on the transport studies- no direct magnetic measurements were performed [3, 4, 7]. To overcome the limitations of the conventional molecular spintronics devices magnetic tunnel junctions (MTJ), produced by sandwiching an insulator between two ferromagnetic electrodes, were utilized as the test bed [5, 6, 8]. Under this approach an MTJ with the exposed side edges can enables the covalent bonding of desired molecular channels onto the two ferromagnetic electrodes along the junction perimeter [5]. These molecules can be single molecular magnets [9], porphyrin [10], single ion molecules, and DNA [11]. This approach is equally capable of utilizing alkane like simple molecules with low spin orbit coupling and Zeeman splitting. The MTJ based molecular spintronics device (MTJMSD) approach enabled us to study the impact of paramagnetic molecules on the spin transport and magnetic properties of MTJs. This paper discusses experimental studies conducted before and after transforming an MTJ into MTJMSD. We report the observation of paramagnetic molecule induced dramatic changes in spin transport of an MTJ. We also report complementary SQUID and magnetic resonance studies exploring the underlying mechanism behind the impact of covalently bonded molecular channels between two ferromagnets of an MTJ testbed.

**Experimental details:** To produce MTJ for MTJMSD, a bilayer ferromagnetic electrode was deposited by sequentially depositing ~ 3 nm tantalum (Ta), 5-7 nm cobalt (Co) and 5-3 nm NiFe with 79% Ni by weight. In the next step photolithography was performed to create a cavity with vertical side edges (Fig. 1b). This cavity helped defining the lateral dimensions of the deposition of a 2 nm thick alumina (AlOx) (Fig. 1c), and top electrode comprising a ~7 nm thick NiFe and ~3 nm tantalum (Ta) (Fig. 1d). The deposition of AlOx and top electrode via the same photoresist (PR) cavity ensured that along the MTJ edges the minimum gap between the two ferromagnetic electrodes is equal to the AlOx insulator thickness (Fig. 1a and g). The liftoff of PR produced Ta/Co/NiFe/AlOx/NiFe/Ta MTJ (Fig. 1d). This MTJ possessed exposed side edges (Fig. 1f). Along the exposed side edges organometallic molecular complexes (OMCs) [12] were bridged across the AlOx to complete the MTJMSD fabrication. These OMCs exhibited S=6 spin state in the bulk powder form at <10 K [12]. An OMC possessed cyanide-bridged octametallic molecular cluster, $[(pzTp)Fe^{III}(CN)_3]_4[Ni^{II}(L)]_4[O_3SCF_3]_4$ [(pzTp) = tetra(pyrazol-1-yl)borate; L = 1-S(acetyl)tris(pyrazolyl)decane][12] chemical structure. With the help of thiol functional groups, an array of OMCs were covalently-linked onto the NiFe layer of the top and bottom electrodes (Fig.1f). The process details for attaching OMCs and the fabrication of tunnel junction with the exposed side edges are published elsewhere [13, 14]. NiFe film was extensively employed in MTJMSDs. NiFe electrode possessed many useful attribute to enable the fabrication of an MTJMSD [15, 16]. As shown in the reflectance study on the NiFe surface, NiFe is stable in the ambient conditions and start oxidizing upon heating around 90 ºC (Fig. 1g). We also tested the efficacy of liftoff based molecular device fabrication by attaching and detaching the OMC between gold (Au) and nickel (Ni) electrode (Fig. 1h). For the magnetic studies cylindrical MTJs were produced by the utilization of photoresist cavities (Fig. 1i). After the thin film deposition and liftoff of photoresist we produced an array of ~7000 MTJ cylindrical pillars (Fig. 1j). Subsequently, MTJs were transformed into MTJMSDs. SQUID and ferromagnetic resonance were performed as per the details furnished elsewhere [8].

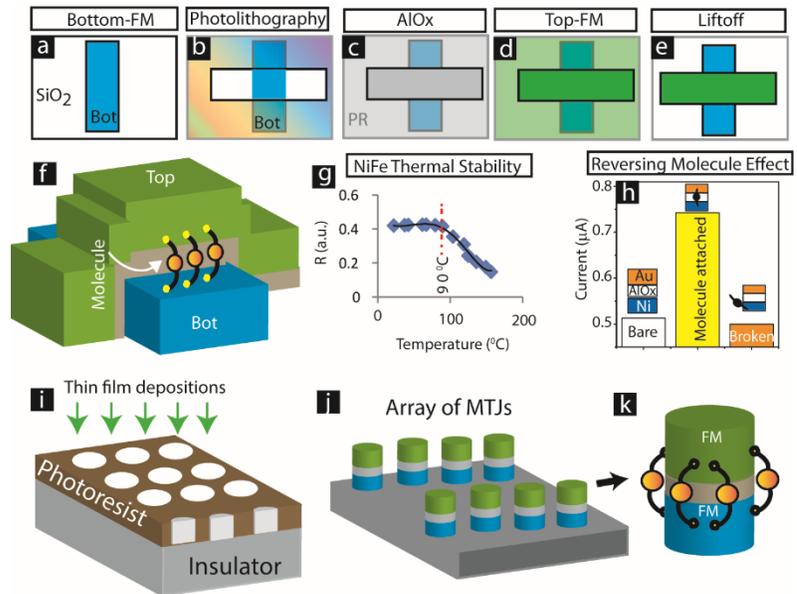

Fig.1: MTJMSD fabrication steps: (a) deposit the bottom ferromagnetic (bottom-FM) electrode on the insulating substrate, (b) creates photoresist (PR) cavity pattern for the deposition of (c) ~2 nm AlOx and (d) top ferromagnetic electrode (top-FM). (e) Liftoff step produces an MTJ with the exposed sides where (f) OMC molecules are chemically bridged across the AlOx insulator to produce an MTJMSD. (g) NiFe films is determined by reflectivity study. (h) OMC were connected and disconnected on a tunnel junction to retain bare tunnel junction transport attributes. Fabrication steps for magnetic studies of MTJMSD. (i) In ~7000 photoresist cavities thin films were deposited to mass produce (j)an array of cylindrical MTJs with exposed edges. (k) Individual MTJ cylinders were transformed into MTJMSD by attaching OMCs.

**Results and discussion:** It must be noted that we formed covalent bonding between the OMCs and the NiFe ferromagnet using thiol functional groups. Thiol anchoring group are known for establishing strong coupling with the metal electrodes in molecular devices [7]. Simple thiol assembly on non-magnetic gold surface was effective in developing permanent magnetism [17]. Also, self-assembly of molecules with thiols end group has enhanced environmental stability and corrosion resistance of ferromagnets[18]. We noted that OMCs interfacing with two ferromagnets of an MTJ influenced the magnetic properties of the ferromagnetic electrodes [8]. Recent, theories developed for interface based devices confirmed that molecule can significantly impact density of states of the ferromagnets [19, 20]. Our prior experimental and theoretical studies [8] also indicated that MTJMSD discussed here are inclined to develop strong coupling between thiol-terminated OMC and NiFe ferromagnetic electrodes.

To study the impact of OMC we conducted current-voltage (I-V) studies on MTJ testbed before and after transforming them into MTJMSDs. Generally, OMCs lowered the current below the bare MTJ's leakage current level via the ~ 2 nm AlOx tunneling barrier (Fig. 2a). In the specific case discussed here, OMCs lowered the current by one order for the Ta/Co/NiFe/AlOx/NiFe/Ta

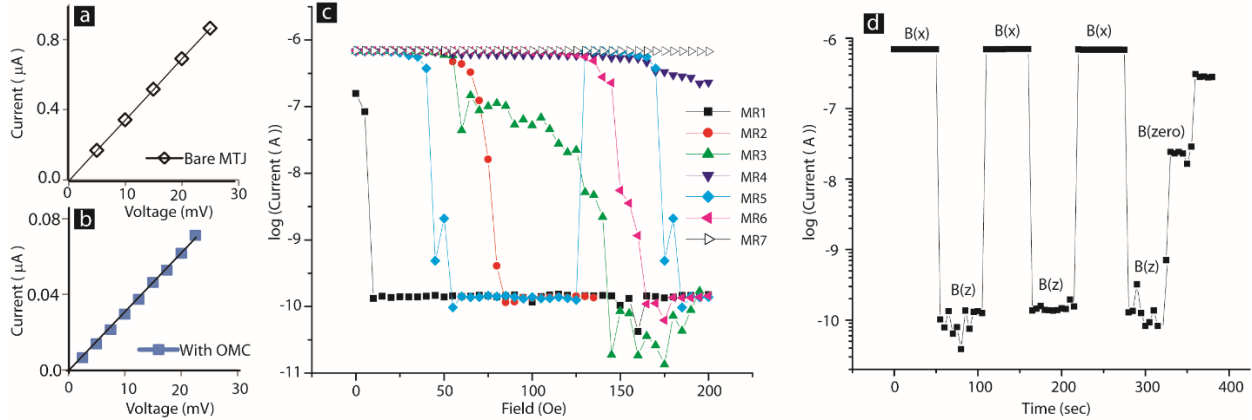

Fig. 2: Current-voltage study of the MTJ (a) before and after attaching (b) OMCs across the tunnel barrier along the exposed edges. (c) Magneto-resistance (MR) studies on MTJMSD showing several orders of change current at 50 mV bias and under in-plane magnetic field. (d) Effect of changing magnetic field direction from inplane to out of the plane on MTJMSD.

MTJ (Fig. 2b). However, after multiple current –voltage studies overall MTJMSD current settled in the high current state. Higher current state also appeared after magneto-resistance (MR) studies. The magnitude of high current state could be one to two orders of magnitude higher than the leakage current via the ~ 2 nm AlOx tunneling barrier. We performed MR studies to investigate the effect of magnetic field on the MTJMSD transport. We observed that application of in plane magnetic field was effective in bringing MTJMSD into suppressed current state by several orders (Fig. 2c). However, it appears that application of 50 mV voltage in MR studies promoted the higher current state MTJMSD (Fig. 2c). Hence, during MR studies a competition occurred between voltage and magnetic field to control the MTJMSD current state. In the first MR study (MR1), the application of magnetic field switched MTJMSD's current state by more than three orders (Fig. 2c). In the second MR study (MR2) higher current remained stable for up to ~170 Oe and then switched to lower current state (Fig. 2c). In 3$^{rd}$ MR study (MR3) current reduction was not sudden or sharp. MTJMSD transitioned through several current values with increasing magnetic field (Fig. 2c). The fourth MR study (MR4) did not show complete transition from sub µA to fully suppressed current state (Fig. 2c). It appears that prolong voltage application promoted higher current state. In the sixth MR study (MR6) MTJMSD lowered to suppress current state and then sprung back to higher current state. Finally, during the seventh MR study (MR7) MTJMSD settled in the higher current state. This behavior was observed on multiple devices. Another device exhibiting this behavior is shown in the supplementary material (Fig. 1S). Even though similar pattern repeated on multiple samples, but switching fields were not consistent over multiple MR studies.

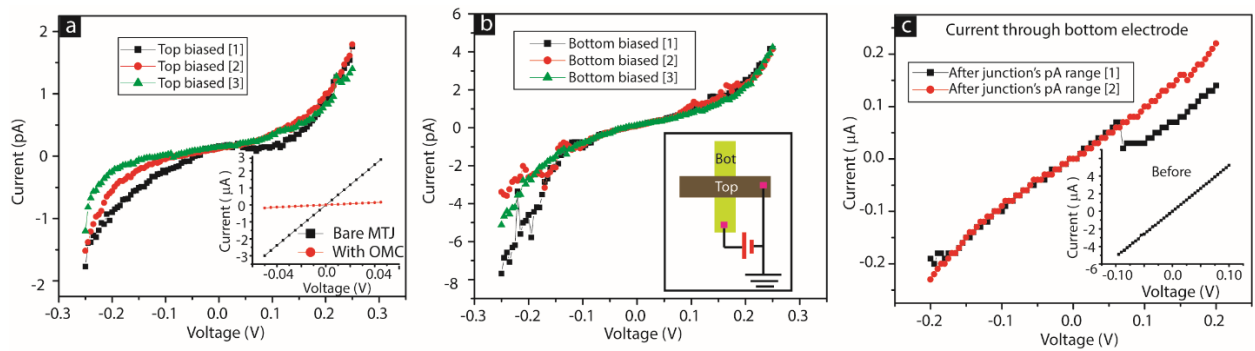

Fig. 3 (a) Stabilized pA level current state on MTJMSD. Inset of this panel shows bare tunnel junction current before and after hosting OMC molecular channels. (b) Effect of changing biasing lead on pA level current state. (c) Current-voltage study through bottom electrode before and after the stabilization of pA level current state.

We also observed that the application of alternating inplane and out of the plane magnetic field was effective in switching MTJMSD between high and low current state (Fig. 2d). The application of 200 Oe magnetic fields was applied inplane and out of the plane. MTJMSD was subjected to 50 mV bias during this study. Switching off magnetic field appeared to stabilize the higher current state (Fig. 2d).

We further investigated the suppressed current state. It is also noteworthy that due to the instrument limitation we were not able to capture exact transport properties of the MTJMSD in the suppressed current state during MR studies discussed in figure 2c-d. The observed sub nA level suppressed current state during MR study was more due to the instrument measurement limit. For the study of suppressed current state a Kithley 6430 source meter with fA level measurement capability was utilized. For this study another sample with exact same configuration was prepared in the different batch. This sample also showed consistent behavior and MTJ testbed current reduced below the leakage level after bridging of the OMC channels (Fig. 3a inset). This behavior matched with most of the samples we studied and two other samples (Fig. 2 and Fig 1S in the supplementary material) discussed in this paper. Knowing that magnetic field produced suppressed current state we applied magnetic field by using a permanent magnet producing ~1000 Oe inplane magnetic field in MTJMSD vicinity. As expected MTJMSD current reduced and settled in sub pA range at room temperature (Fig. 3a). This suppressed current state was persistent over multiple I-V studies. Flipping the biasing direction from top electrode to bottom electrode produced minor impact on transport (Fig. 3b). Results from three consecutive I-Vs are shown in Fig. 3a and Fig. 3b. However, after leaving the sample intact for some time brought the MTJMSD back in the higher current state. To ensure that electrodes to the junction were intact the transport through the top and bottom electrodes were studied before and after setting MTJMSD into suppressed current state. Current through top electrode was almost the same before and after the setting of suppressed current state. However, the bottom electrode appeared to show reduced current state, but it was still almost six orders higher than the magnitude of the observed suppressed current state (Fig. 3c).

It is highly intriguing that ~10,000 OMCs along the MTJ edges produced a suppressed current state below the leakage current level through the planar area (Fig. 2-3). This observation of OMC induced current suppression was only observed with the magnetic electrodes. On the other hand, nonmagnetic tunnel junctions, prepared with gold, copper, and tantalum electrodes, exhibited an increase in current due to the addition of OMCs channels along the side edges. The key difference in the magnetic and nonmagnetic electrode is the unequal density of states for the spin up and spin down electrons. Spin transport through a MTJ depends on the spin density of states of the ferromagnetic electrodes. A plausible reason behind the current suppression may be due to OMC induced changes in magnetic properties or spin density of states of the ferromagnetic electrodes.

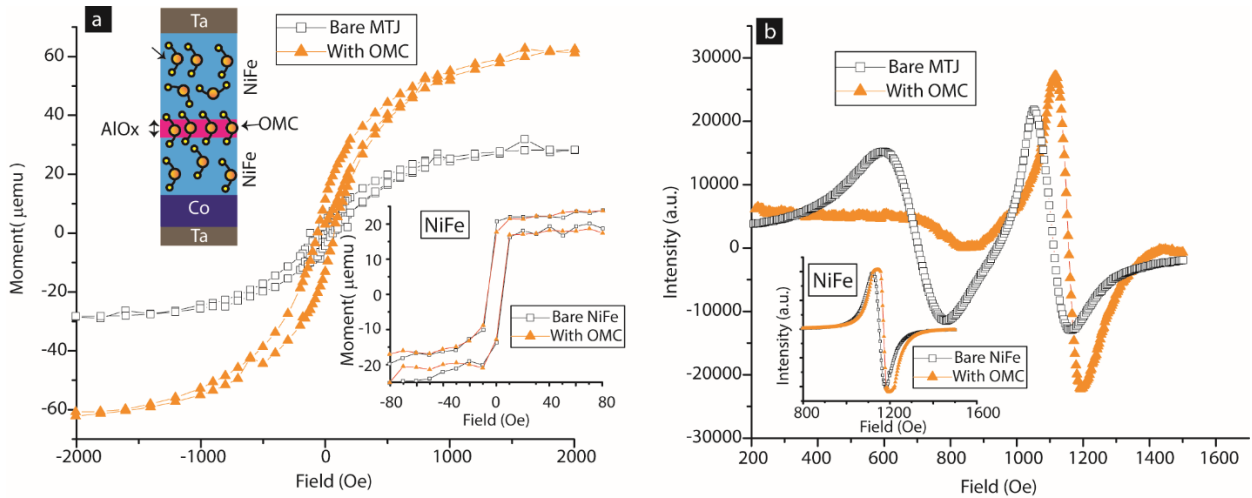

Fig. 4 Magnetic studies of 7000 MTJMSD by SQUID and FMR. (a) SQUID magnetometer study of MTJ before and after hosting OMC channels across AlOx insulator. Top left inset shows that OMC can bridge across AlOx and also bond with the exposed magnetic layers, mainly NiFe. Bottom right inset show magnetic study of NiFe before and after interacting with OMCs. (b) FMR study of MTJ before and after hosting OMC channels. Inset graph shows the impact of OMC on NiFe film only.

In the prior study, we have provided insight regarding OMC induced impact on inter-ferromagnetic electrode coupling and magnetic ordering of the ferromagnetic electrode itself [8]. This prior work presented experimental SQUID, Ferromagnetic resonance (FMR), and magnetic force microscopy studies and it also presented theoretical insight from Monte Carlo simulations [8]. However, the top electrode in the current study differed with previous study. In the current study top electrode is made up of ~ 7 nm NiFe with ~ 3 nm Ta on top. In the previous study top electrode was ~10 nm thick NiFe. To investigate direct correlation between magnetic properties and the transport studies discussed in Fig. 2 and 3 separate SQUID magnetometer and FMR studies were conducted (Fig. 4). These studies were performed on an array of ~ 7000 cylindrical MTJs before and after treating them with OMCs. The fabrication process for the cylindrical MTJ is shown in Fig. 1i-k. The experimental details for SQUID magnetometer and FMR studies are discussed elsewhere [8]. SQUID study showed that the magnetic moment of the MTJ (TaCoNiFe/AlOx/NiFe/Ta) almost doubled after hosting OMC molecular channels (Fig. 4a). This is an average effect from 7000 cylindrical MTJs and clearly suggests that OMCs dramatically affected the magnetic moment of MTJ. This study also confirmed that OMC impact is strong and can spread all over the microscopic junction. The mechanism behind doubling of magnetic moment is not clear to us. However, it is quite well known that a molecule with net spin state can enable spin filtering and that may significantly enhance the degree of spin polarization. For instance if spin polarization for ferromagnet changes from ~0.5 to 1 then magnetic moment for each electrode may double to produce observed enhancement (Fig. 4a).

It is noteworthy that only a small fraction of OMCs formed bridge across insulator. Most of the OMCs stick to the NiFe ferromagnetic electrode regions away from the tunnel barrier (top left inset of the Fig. 4a). SQUID magnetometer study was conducted before and after treating unpatterned NiFe film with OMCs. SQUID magnetometer study did not produce any noticeable change due to OMC (bottom right inset Fig.4a). This study also confirmed that magnetometer performed accurate measurement. This study also suggested that OMCs that matters in an MTJMSD are those which bridged across insulating AlOx. We hypothesized that other magnetic measurements must be able to record such a strong OMC induced change in MTJ properties. We performed ferromagnetic resonance (FMR) at room temperature and found that OMC treated sample with ~7000 MTJs was starkly different from the untreated MTJ. FMR of the OMC treated sample showed one pronounced resonance mode around ~1200 Oe. However, bare MTJ produced an optical mode (lower intensity mode) around ~650 Oe and acoustic mode (higher intensity mode) around 1100 Oe. According to FMR theory of coupled ferromagnetic films the bare MTJ seems to possess ferromagnetic coupling [21]. Luckily, prior theoretical studies have explored the FMR spectra from analogous system. The development of very strong coupling between the ferromagnetic electrodes produced one resonance mode [22]. Under strong coupling limit two ferromagnetic electrodes behave like one ferromagnetic electrode [22]. Similar to SQUID magnetometer study we also investigated the FMR response from the NiFe. OMC treatment did not produce significant change (inset Fig.4b). Similar to SQUID magnetometer study, FMR study also confirmed that OMC bridges strongly influenced the magnetic coupling between two ferromagnetic electrodes of the MTJ and dominated the spin transport.

At this point we are unsure about why the application of magnetic field produced the suppressed current state on MTJMSD (Fig. 2-3). We hypothesize that the magnetic field play a role in configuring the large mass of the ferromagnetic electrodes in accordance with the molecule influenced area. It is noteworthy that overall magnetic leads are nearly one cm long and hence whole length of ferromagnetic leads cannot be influenced by the molecular junction. The application of magnetic field may optimally aligned magnetic moment of the ferromagnetic electrode. Disciplining the large mass of ferromagnet may promote or restrict the transport of spins through the highly ordered OMC affected junction. We also observed that repeating I-V studies or prolong application of external voltage promoted the higher current state on an MTJMSD. We hypothesize that the application of voltage assist in transporting unpolarized electrons from the bulk of the leads into the OMC affected regions and hence, promoting disorder to yield settlement of higher current state. Pinpointing exact mechanism is yet to be done and will require first principle theoretical studies and extensive magnetic studies.

Based on the I-V study we estimated the effective barrier heights and barrier thicknesses using Brinkman tunneling models [23]. A bare MTJ exhibited ~2.2 nm barrier thickness and ~0.7 eV barrier height. After hosting OMC channels along the exposed side edges an MTJ become MTJMSD and showed very different barrier properties in the ~micro amp range high current state and pA range suppressed current state. According to modelling results in the µA range high current state an MTJMSD exhibited ~ 1.2 nm barrier thicknesses and ~0.4 eV barrier height. The same MTJMSD in the pA level suppressed current state (Fig. 3) exhibited ~1.4 nm barrier thickness and ~ 2 eV barrier heights. Importantly the barrier thickness after hosting OMCs is equivalent to the length of a decane molecule chain that connected the core of OMC molecules to the NiFe electrode in an MTJMSD (Fig. 6a).

We investigated the effective density of states before and after transforming MTJ into an MTJMSD. According to prior literature density of state in the molecule affected regions depends on barrier height ($\phi$) and the molecular coupling strength ($\omega$) [19]. One can calculate the effective density of states ($D(E)$) the following equation:

$$D(E) = \frac{\omega/2\pi}{(\phi)^2 + (\frac{\omega}{2})^2} \qquad (1)$$

We estimated the relative increase in the strength of exchange coupling by utilizing the following equation [24].

$$\omega = J_0 \exp(-\beta.d) \qquad (2)$$

According to the analysis of charge transport data, barrier thickness ($d$) decreased from ~2.2 nm to ~1.2 nm range. This effective barrier thickness of MTJMSD corresponded to alkane tether length of ~1.2 nm. The MTJMSD barrier thickness was comparable in the high and low current state. The barrier height ($\phi$) can be utilized to obtain tunneling decay factor $\beta$ according to the relation $\beta = 10.2$ nm$^{-1}\sqrt{(\phi/\text{eV})}$ [25]. In the equation (2) $J_0$ correspond the coupling energy between two magnetic atoms that decreased exponentially with barrier thickness and $\beta$. According to literature Ni-Ni atoms coupling energy is ~12 µeV [26]. It is noteworthy that in this simplified analysis we have not incorporated spin fluctuation that may dramatically enhance the exchange coupling [2, 27]. The density of the state of bare MTJ was calculated to be ~$10^{-9}$ states/eV for barrier thickness 2.2 nm and 0.7 eV barrier heights. The MTJMSD's density of states in the high current state was ~$10^{-7}$ states/eV. However, the density of states in suppressed current state was ~$10^{-13}$ states/eV. It is noteworthy that spin polarization has been routinely calculated by the high and low current states for the magnetic electrodes in magnetic tunnel junctions [28-30]. Nearly six orders of magnitude change in the high and low current states suggests that OMC have produced nearly 100% spin polarized magnetic electrodes. We believe that OMC's spin state played a crucial role in spin filtering [31]. The OMC molecule is capable of attaining the S=6 spin state in the isolated state [12, 32] . We presumed that this S=6 state is still applicable when an OMC is bonded to the ferromagnetic electrodes (Fig. 6a). In this state, it seems that iron (Fe) and Ni atoms of the OMC clusters are open for accommodating only one specific type of spin due to selection rule and leading to spin filtering [7].We also hypothesize that method of chemically bonding OMC to the magnetic electrode also played crucial role. Several thousands of OMCs are connected to two ferromagnetic electrodes via alkane barrier and thiol chemical bonding (Fig. 6a). We utilized thiol (-S) bonds to form covalent bonding between OMC and NiFe to produce very strong molecule induced exchange coupling. Prior studies illustrated the strong impact of thiolated molecules on the magnetic and transport properties of metallic electrodes [7, 17, 33].

Our magnetic studies provide clear evidences that OMC produced transformative exchange coupling on an MTJ [8] with similar thin film configuration as discussed here. Our experimental

FMR study also confirmed that OMC have dramatically enhanced the exchange coupling between top and bottom ferromagnetic electrodes (Fig. 5b). We noted that adding Ta on top of the previously studied MTJ has been helpful in observing current switching in the high and low current state of the MTJMSD. However, we also noted that FMR response from the MTJ with Ta (Fig. 5b) and without Ta on top [8] were very different. The further exploration about Ta role will be part of future study and beyond the scope of this paper.

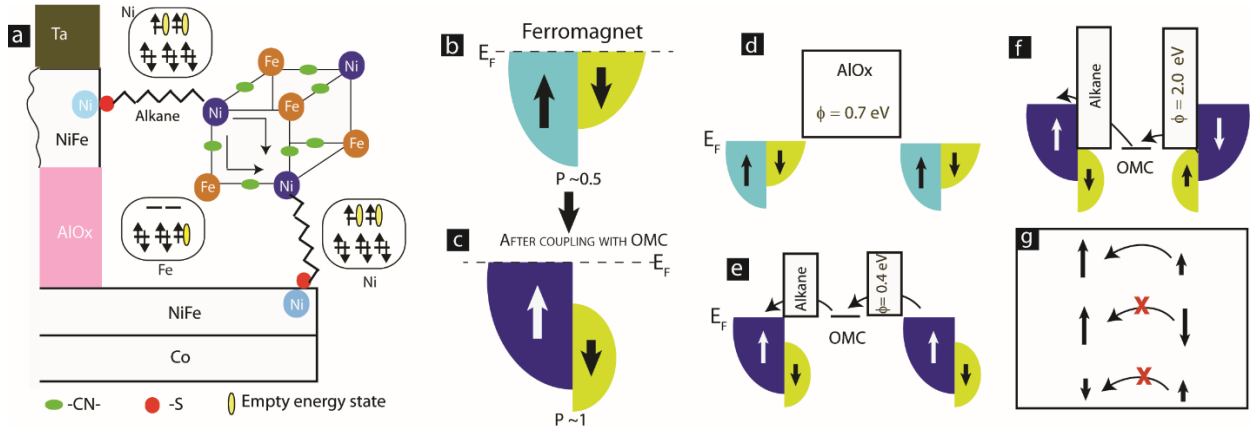

Fig. 6: (a) Schematic of an OMC chemically bonded with two NiFe layers along the MTJ edge. Simplified molecular structure shows energy levels of the Fe and Ni atoms. Density of states of a ferromagnetic electrode (b) before, and (c) after affected by the molecular channels. (d) Tunneling barrier between two ferromagnetic layers for a bare magnetic tunnel junction. OMC provide intermediate quantum well like energy levels for the spin transport. The conceptual depiction of tunneling barrier between ferromagnet and OMC energy level in the (e) high current state and (f) suppressed current state. (g) Possible spin pathways in the suppressed current state.

We conjectured that strong OMC based coupling impacted the whole microscopic junctions and changed the spin polarization of the ferromagnetic electrodes. Recent studies shine light on molecule impact on the spin density of states on ferromagnets near interface regions [20]. According to our SQUID magnetometer study on an array of ~7000 MTJ OMC showed almost 100% increase in the magnetic moment (Fig. 5a). OMCs seem to change the spin density of states of the ferromagnetic electrodes. The spin density of state before OMC interaction (Fig. 6b) is presumably adjusted to new configuration (Fig. 6c) leading to high spin polarization. The overall tunneling barrier energy band diagram has changed from AlOx dominant (Fig. 6d) to OMC dominant (Fig. 6e-f). We hypothesize that to establish high current state majority spin density of states of the two ferromagnetic electrodes is parallel to each other (Fig. 6e). The effective barrier height in this state is ~0.4 eV, as calculated from the experimental result (Fig. 6e). However, in the suppressed current state we hypothesize that spin transport depends on the net energy barrier with respect to the minority spin density of states on an MTJMSD (Fig. 6f). In this state the majority and minority spins are antiparallel to each other and hence do not allow transport. In this state transport between minority and majority spin is possible (Fig. 6g). Due to selection rule spin from majority band cannot tunnel to antiparallel majority band of the other magnetic electrode (Fig. 6g). Similarly, the spin from minority band of one magnetic electrode cannot tunnel to antiparallel minority band of the other magnetic electrode (Fig. 6g).

**Conclusions:** We discussed the observation of molecule induced dramatic changes in the magnetic and transport properties of the magnetic tunnel junctions. OMCs were chemically bonded to ferromagnetic electrodes to bridge them across the insulating spacer along the exposed edges. These molecules dominated the MTJMSD's spin transport and produced several orders gap in the high and low current state at room temperature. The estimation of transport data in the low and high current states indicated that molecules changed the spin density of states. Potential reason behind modification in the density of states is the molecule induced spin filtering effect. Transport studies are strongly supported by the experimental magnetic studies. It is also noteworthy that our experimental studies provide a platform to connect a vast variety of ferromagnetic leads to the broader array of high potential molecules such as single molecular magnets [1], porphyrin [10], and single ion molecules [34] etc. The strength of exchange coupling between ferromagnetic electrodes and molecules can be tailored by utilizing different tethers and terminal functional groups [7]. The MTJMSD can provide advanced computer devices by serving as a testbed for the molecule based quantum computation devices [34-37].

**Acknowledgments:** Pawan Tyagi thanks Dr. Bruce Hinds and Department of Chemical and Materials engineering at University of Kentucky for facilitating experimental work on MTJMSD during his PhD. OMC were produced Dr. Stephen Holmes's group. The preparation of this paper and supporting studies were in part supported by National Science Foundation-Award (Contract # HRD-1238802), Department of Energy/National Nuclear Security Agency (Subaward No. 0007701-1000043016), and Air Force Office of Sponsored Research(Award #FA9550-13-1-0152). Any opinions, findings, and conclusions expressed in this paper are those of the author(s) and do not necessarily reflect the views of any funding agency and authors' affiliations.